# Optical module HEW simulations for the X-ray telescopes SIMBOL-X, EDGE and XEUS

D. Spiga*

INAF/ Osservatorio Astronomico di Brera, Via E. Bianchi 46, I-23807 Merate (LC), Italy

**ABSTRACT**

One of the most important parameters defining the angular resolution of an X-ray optical module is its Half-Energy Width (HEW) as a function of the photon energy. Future X-ray telescopes with imaging capabilities (SIMBOL-X, Constellation-X, NeXT, EDGE, XEUS,…) should be characterized by a very good angular resolution in soft (< 10 keV) and hard (> 10 keV) X-rays. As a consequence, an important point in the optics development for these telescopes is the simulation of the achievable HEW for a system of X-ray mirrors. This parameter depends on the single mirror profile and nesting accuracy, but also on the mirrors surface microroughness that causes X-ray Scattering (XRS). In particular, owing to its dependence on the photon energy, XRS can dominate the profile errors in hard X-rays: thus, its impact has to be accurately evaluated in every single case, in order to formulate surface finishing requirements for X-ray mirrors. In this work we provide with some simulations of the XRS term of the HEW for some future soft and hard X-ray telescopes.

## 1. INTRODUCTION

Future X-ray telescopes, like SIMBOL-X[1], Constellation-X/HXT[2], NeXT[3], and XEUS/HXT[4], will extend the imaging capabilities of present soft X-ray telescopes (Chandra, Newton-XMM, Swift-XRT) to the hard X-ray band (E > 10 keV). A basic feature of these instruments will be a noticeable area in hard X-rays, that will allow a large sensitivity gain and the consequent chance of studying fainter cosmic X-ray sources beyond 10 keV. In the case of SIMBOL-X, this leap forward will be made possible by the adoption of a *very long (20 m) focal length*, which enables the utilization of very shallow incidence angles (0.11 ÷ 0.25 deg) without a reduction of the mirror module size, and, consequently, of the collecting area; such a focal length can be managed flying the optics and the focal plane instruments on two separate spacecrafts in *formation-flight* configuration[5]. Moreover, graded multilayer coatings[6,7] for X-ray mirrors will be utilized to extend the reflectivity of the optics to a wide X-ray energy band, up to 80 keV. In the case of XEUS, these techniques will be used in conjunction with the adoption of an optical module with a very large diameter (> 4 m). With incidence angles between 0.27 and 0.86 deg, this implies a 35 m focal length. Other X-ray telescopes, like EDGE[8], with a much shorter focal length (2.75 m) will instead take the advantage of a very wide Field of View (up to 1.5 deg diameter for the Wide Field Imager instrument) in the soft X-ray energy band (0.3 ÷ 6 keV). In this case, it will be the choice of X-ray mirrors with a properly-optimized polynomial[9] design that will enable a good imaging quality also for off-axis sources[10].

For the observation of very distant X-ray sources, often faint and to be resolved in crowded fields, a *very good angular resolution* is an essential requirement for all these X-ray telescopes. A suitable parameter to express the optical performance of X-ray optics is the HEW (*Half-Energy Width*, the angular diameter collecting 50% of the reflected photons). In the angular resolution determination, the nominal optic profile plays an essential role: Wolter-I mirrors (SIMBOL-X) return a very good on-axis HEW, but it gets degraded as a source moves off-axis. Polynomial mirrors are specifically conceived to return a good HEW over a large Field of View, at the expense of a slight degradation of the on-axis imaging quality. Finally, double-cone optics are a suitable solution to construct very large diameter optics, even though the achievable imaging quality is usually worse than with the profiles mentioned above. It is easily understood that deviation of mirrors from nominal profile profiles, and mirrors misalignment, degrade the focusing sharpness and increase the HEW. This geometrical effect on the HEW is independent of the photon energy.

---

\* e-mail: daniele.spiga@brera.inaf.it, phone +390399991146, fax +39039999160

Another very important factor that should be accounted for is the *X-ray scattering (XRS)*. This effect is caused by the *microroughness* of reflecting surfaces, which is often quantified by means of the surface rms, σ. The XRS intensity is an increasing function of the photon energy: therefore, if the mirrors surface is not very smooth, this problem can even dominate the mirror profile errors at high energies. Fortunately, a well consolidated theory of X-ray scattering from rough surfaces[11] allow us to evaluate its impact on the HEW as the photon energy increase. One of the most useful results is that the intensity of XRS is proportional to the *Power Spectral Density* (*PSD*) of the surface roughness[12], i.e., the distribution of roughness at various spatial frequencies *f*. Moreover, the incidence angle $\theta_i$, the scattering angle $\theta_s$ and the photon wavelength λ are related to the spatial frequency of the PSD by the simple relation:

$$f = \frac{\cos\vartheta_i - \cos\vartheta_s}{\lambda}. \tag{1}$$

From this simple, classical formula it becomes apparent that large *f* (or equivalently, small spatial wavelengths $\ell = 1/f$) are associated to scattering at large $\theta_s$: for this reason, a PSD with a large high frequency content may be responsible for a serious degradation of the HEW at high energies. We could not reach this conclusion only from the knowledge of the rms roughness σ, that can be computed from the integration of a PSD *P(f)* over a specific frequency range Δ*f*,

$$\sigma^2 = \int_{\Delta f} P(f)df. \tag{2}$$

In fact, the σ parameter of an X-ray mirror can neither suffice to provide us with a complete characterization of its surface finishing level, not it enables the evaluation of imaging degradation due to XRS in a X-ray telescope, as a function of λ. Conversely, in this work we will firstly see (Sect. 2) that information concerning the PSD of the mirrors surface can be immediately utilized to determine and derive the XRS contribution to the HEW in an assigned photon energy range.

To this end, we will make use of an analytical formalism[13], also introduced and discussed with more details in another paper of the present proceedings volume[14]. In Sect. 3 we will make use of those results to simulate the HEW degradation due to X-ray scattering for the SIMBOL-X mirrors from 0.5 to 80 keV: a comparison of results with the HEW scientific requirements will provide us with a *possible* formulation of microroughness tolerances, in terms of PSD, for the SIMBOL-X mirrors. In Sect. 4 we extend this method to the problematic of the EDGE-*Wide Field Imager* optical module: even if EDGE will operate in soft X-rays (hence, XRS could seem a minor issue at first glance), a quantification of the X-ray scattering has to be carried out to establish the surface smoothness needed to keep the HEW close to 15 arcsec. In Sect. 5 we perform the same analysis for the XEUS X-ray telescope, with reference to the present baseline[15], that foresees the implementation of *Silicon pore optics*[16]. Some basic features of the mentioned telescopes are reported in Tab. 1. The results are briefly discussed in Sect. 6.

Tab. 1: main features of the optical modules of the SIMBOL-X, EDGE-WFI, XEUS X-ray telescopes

|  | **SIMBOL-X** | **EDGE-WFI** | **XEUS** |
| --- | --- | --- | --- |
| **Focal length** | 20 m | 2.75 m | 35 m |
| **Min diameter** | 300 mm | 305 mm | 1.3 m |
| **Max diameter** | 700 mm | 700 mm | 4.2 m |
| **Min incidence angle** | 0.11 deg | 0.79 deg | 0.27 deg |
| **Max incidence angle** | 0.25 deg | 1.81 deg | 0.86 deg |
| **Energy band** | 0.5 – 80 keV | 0.3 – 6.0 keV | 0.1 – 40 keV |
| **Number of shells** | 100 | 43 | 2200 |
| **Effective area (1 keV)** | ~ 1400 cm² | 580 cm² | 5 m² |
| **Effective area (30 keV)** | ~ 450 cm² | -- | 400 cm² |
| **Field of View** | 12 arcmin | 1.5 deg | 7 arcmin (WFI) |
| **Required HEW** | 15 arcsec (1 keV)<br>20 arcsec (30 keV) | 15 arcsec | 5 arcsec |

## 2. EVALUATION OF XRS IMPACT ON THE HEW FOR X-RAY MIRROR MODULES

Among several methods that can be used to compute the HEW from the surface PSD, we will hereafter adopt a novel method, based of the well-known of theory the XRS from rough surfaces[11] *to perform a direct translation of a surface PSD into a H(λ) function in a wide energetic range.* Here we recall only the final results for the case of a *double reflection* mirror shell: if $\theta_i$ indicates the incidence angle, $P(f)$ the mirror surface PSD, the HEW scattering term $H(\lambda)$ can be computed by deriving $f_0$ from the integral equation[13]

$$\int_{f_0}^{2/\lambda} P(f)df = \frac{\lambda^2 \ln(4/3)}{16\pi^2 \sin^2 \vartheta_i}, \tag{3}$$

and substituting $f_0$ in the Eq. 1, in the following approximated form:

$$H(\lambda) = \frac{2\lambda f_0}{\sin \vartheta_i}. \tag{4}$$

The factor ln(4/3) is related to the choice of a double reflection mirror. The results of these equations can be compared with ray-tracing numerical routines, properly modified in order to allow for X-ray scattering due to surface roughness[14]. The Eq. 3 can be also reversed, enabling the conversion of a given $H(\lambda)$ function into the corresponding $P(f_0)$,

$$P(f_0) = -\frac{\lambda \ln(4/3)}{4\pi^2 \sin^3 \vartheta_i} \left[ \frac{d}{d\lambda}\left(\frac{H(\lambda)}{\lambda}\right) \right]^{-1}. \tag{5}$$

If the PSD well approximates a *power-law*[17] model $P(f) = K_n / f^n$ ($1 < n < 3$), we can derive that *also the H(λ) function is a power-law*, and the spectral index of $H(\lambda)$ is related to that of the PSD:

$$H(\lambda) = 2 \left[ \frac{16\pi^2 K_n}{(n-1)\ln(4/3)} \right]^{\frac{1}{n-1}} \left( \frac{\sin \vartheta_i}{\lambda} \right)^{\frac{3-n}{n-1}} \tag{6}$$

for a extremely steep (n ~ 3) PSD, the HEW becomes a *constant*. On the contrary, for a smoother PSD (n ~ 1) the HEW increases with the photon energy very steeply. *This makes apparent the importance to obtain not only low, but also steep surface PSDs for X-ray mirrors.* The influence of the spectral index will also be discussed by an example in Sect. 5.

Once computed the $H(\lambda)$ energy-dependent term of the HEW, the total HEW($\lambda$) should arguably be obtained by adding in quadrature the figure error contribution to the HEW, $H_0$:

$$HEW^2(\lambda) = H_0^2 + H^2(\lambda). \tag{7}$$

Usually X-ray optical modules comprise several confocal mirror shells. Once computed the HEW of each mirror shell, the HEW of the optical module can be *approximately* obtained by averaging the single contributions $HEW_k$ (where $k = 1,2...$ is the mirror shell index) with appropriate weights. At a first-order approximation, the weights equal the mirror shells effective areas $A_k$, times the respective PSFs $S_k$ (normalized to 1) evaluated at the respective Half-Energy-Radii $\Omega_k$. = $HEW_k/2$,

$$HEW_T = \frac{\sum_k HEW_k S_k(\Omega_k) A_{eff,k}}{\sum_k S_k(\Omega_k) A_{eff,k}}. \tag{8}$$

where we omitted the explicit dependences on λ to simplify the notation. The proof of this result is reported in appendix A. We henceforth assume that the variation of the $S_k(\Omega_k)$ factors with $k$ is negligible with respect to that of effective areas $A_k$. Under this last hypothesis, the weights can be simplified, and the Eq. 8 becomes

$$HEW_T \approx \frac{\sum_k HEW_k A_{eff,k}}{\sum_k A_{eff,k}}. \tag{9}$$

If $k = 1$ labels the smallest mirror shell of the optical module (minimum $\theta_i$) and $k = M$ the largest one (maximum $\theta_i$), $A_k$ is an increasing function of $k$ (at a fixed λ). Supposing a regular, decreasing behaviour of $P(f)$, also the $HEW_k$ should be in general an increasing function of $\theta_i$ (Eq. 6), and thereby of $k$. On the contrary, $S_k(\Omega_k)$ decrease as the half-energy radii $\Omega_k$, = $HEW_k/2$ increase, because the PSF is in general a decreasing function of the angular distance from focus. Thus, $S_k(\Omega_k)$ is a decreasing function of $k$. From this, we infer that Eq. 9 assigns larger weights to HEW of mirrors shell with larger size (and larger $\theta_i$) than the more exact Eq. 8, and will also return a larger total $HEW_T$ value.

## 3. HEW SIMULATIONS FOR SIMBOL-X

The very long focal length of SIMBOL-X[1,5] optical module (see Tab. 1) is conceived to allow the incidence angles to take on very small values (from 0.11 to 0.25 deg). From a qualitative application of Eq. 6, we can expect that the adoption of small $\theta_i$ contributes to the reduction of the XRS term of the HEW. However, the energy band of SIMBOL-X is extended over three orders of magnitude, hence the XRS effect will be relevant, if the X-ray mirrors are not very smooth. In this section we will quantify some possible roughness PSD tolerances for SIMBOL-X mirrors.

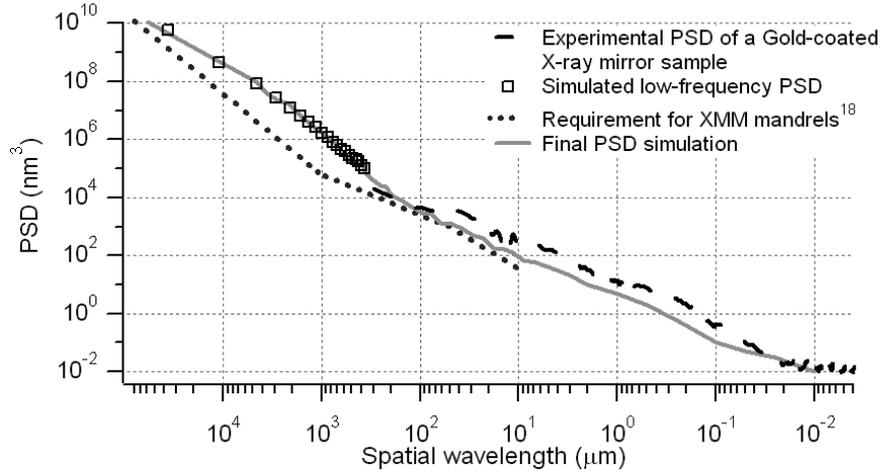

Fig. 1: a realistic PSD for an X-ray mirror can be measured from a Au layer replicated from a Zerodur glass (dashed line). The comparison with requirements of XMM mandrels (dots) shows that even better smoothness level can be achieved between 100 and 10 μm. The grey line represents a possible improvement at high frequencies, consistently with a requirement of 20 arcsec HEW (squares) at 30 keV.

Tab. 2: rms roughness, in selected wavelength windows, for the plotted PSDs in Fig. 1.

| PSD | 1 mm > $l$ > 100 μm | 100 μm > $l$ > 10 μm | 10 μm > $l$ > 1 μm | $l$ < 1 μm |
|---|---|---|---|---|
| **Initial PSD** (squares + dashed line) | 10.2 Å | 3.6 Å | 2.5 Å | 1.8 Å |
| **Improved PSD** (solid line) | 9.3 Å | 2.2 Å | 1.3 Å | 1.0 Å |

The HEW requirement for SIMBOL-X (Tab. 1) is 15 arcsec at 1 keV and 20 arcsec at 30 keV. We can assume that the HEW at 1 keV is essentially due to figure errors, then $H_0$ = 15 arcsec. We may then postulate a slowly-increasing HEW from 15 to 20 arcsec in the energy band 1 ÷ 30 keV and isolate the $H(\lambda)$ contribution by means of Eq. 7. Finally, we may derive a PSD along with Eq. 5 and the respective spatial frequencies with the Eq. 4, adopting for $\theta_i$ the average value of incidence angles of the optics (0.18 deg). *The resulting PSD* (the squares in Fig. 1) *defines a roughness tolerance in the low-frequency regime* ($l$ > 408 μm). In order to extend the PSD, we should extend the HEW specification at higher energies: alternatively, we can use the Eq. 3 with $f_0$ = 1/(408 μm)$^{-1}$ and $E$ = 30 keV to set a maximum allowed value for the integral of the PSD at frequencies larger than $f_0$. *The substitution yields the additional requirement $\sigma(l < 408 \mu m) < 5.5$ Å.*

In order to check the feasibility of such a surface smoothness, we considered the PSD of a smooth X-ray mirror sample, obtained by replicating a Gold layer deposited onto a Zerodur glass. The PSD of this sample has been extensively characterized in the spectral range (300 ÷ 0.01μm) at *INAF-OAB* along with techniques like AFM mapping,

XRS measurements at 8.05 keV, optical profilometry. We can now attempt to connect the low-frequency PSD (squares) with the measured one (dashed line) and adopt the resulting PSD as an initial model. However, the integration of this PSD at spatial wavelengths shorter than 408 μm returns *σ(ℓ < 408 μm) = 6.8 Å*. This indicates that the *high-frequency roughness should still be reduced in order to fulfill the initial HEW requirement*. Interestingly, if we compare the measured PSD with the roughness requirements for the XMM mandrels[18] (dots in Fig. 1), we see that a smoothness level better than the one we just discussed is still conceivable.

Therefore, we considered as a second example a simulated PSD with smaller values at ℓ < 408 μm, in order to satisfy the condition σ(ℓ < 408 μm) < 5.5 Å: also this PSD, that differs from the previous one only at ℓ < 100 μm, is also plotted in Fig. 1. To clarify the difference of roughness levels, we list in Tab. 1 the rms values calculated, from the two considered examples, in selected spatial wavelength windows.

To check the performances expected from the PSDs being addressed here, Eqs. 3 and 4 were used to derive $H(\lambda)$ for every SIMBOL-X mirror shell. A figure error contribution of 15 arcsec was added in quadrature (Eq. 7) to all mirror shells and the final HEW has been obtained by applying the Eq. 9. Notice that a similar calculation was also done in another paper of this volume[14]: unlike that case, however, here we did not set a maximum value for the scattering angles by modifying the upper integration limit of Eq. 3. We allowed instead the HEW to be computed over a several degrees wide angular region, in order to account also for photons scattered at large angles, even if their contribution to the HEW decays quickly as the angular distance from focus becomes large.

The result of the HEW calculation for the two PSDs is shown in Fig. 2. Both HEW trends start from 15 arcsec at 1 keV. After a small increase they reach 18 arcsec at 10 keV, then they diverge from each other. The HEW derived from the "rougher" PSD increase faster and faster, whilst the other one reaches a wide "plateau" at 18 arcsec up to 40 keV, then it restarts to increase, up to 32 arcsec at 70 keV. It is quite clear that only the second PSD fulfills the HEW requirements for SIMBOL-X. This example makes apparent the impact that small differences in roughness at high frequencies can have on the imaging quality at high energies.

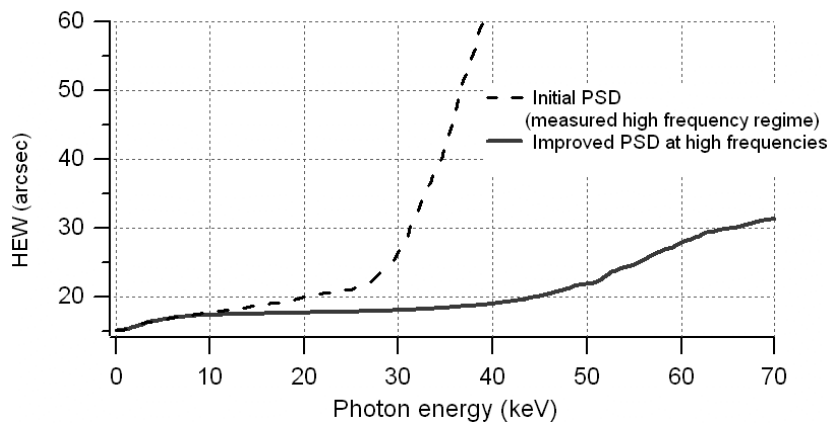

Fig. 2: the HEW trends in the SIMBOL-X energy band as resulting from the weighted sum of the contributions of the single mirror shells, for the two PSDs in Fig. 1. We assumed 15 arcsec figure errors. Only the solid line fulfills the SIMBOL-X HEW requirements. Notice the impact of the high-frequency regime at high energies.

## 4. HEW SIMULATIONS FOR THE EDGE-WFI X-RAY TELESCOPE

With respect to SIMBOL-X, the EDGE[8] X-ray telescope will have a much shorter focal length (2.75 m) and larger field of view (0.7 x 0.7 deg$^2$ with the *Wide Field Spectrometer* and 1.5 deg diameter with the *Wide Field Imager*). As the EDGE sensitivity band is below 6 keV(see Tab. 2), the incidence angles can be larger than for SIMBOL-X. In order to efficiently detect the scientific EDGE-WFI targets an angular resolution of 15 arcsec HEW is required. For instance, we might search for a slowly-increasing HEW with the photon energy from 10 arcsec at 0.3 keV (assumed as figure errors HEW) up to 15 arcsec at 6 keV.

As EDGE will be operated in soft X-rays, XRS might be a less severe issue than for SIMBOL-X. On the other side, the incidence angles are larger than for SIMBOL-X, which implies an X-ray scattering enhancement. A first attempt to set microroughness tolerances can be made assuming as a PSD the dashed line of Fig. 1 (also plotted in Fig. 3): repeating the computation like in the previous section (using Eqs. 3,4,7,8), we obtain the HEW trend shown in Fig. 4 (dashed line). The result is an increasing HEW from 10 at 0.3 keV to 19 arcsec at 5.5 keV. Even if the largest HEW is close to the 15 arcsec requirement, some improvements are possible: hence, in Fig. 3 (dots) we also plot a simulated PSD that returns a HEW which remains under 16 arcsec within the entire EDGE-WFI energy band (Fig. 4, dots). It can be seen from Fig. 3 that the PSD requirement is *less strict than for SIMBOL-X*; this can qualitatively understood by observing that the average value of the $(\sin\theta_i/\lambda)$ ratio is smaller for EDGE-WFI than for SIMBOL-X (see Eq. 6). In Tab. 3 we compare the rms roughness levels of the PSD under consideration in selected windows of spatial wavelengths.

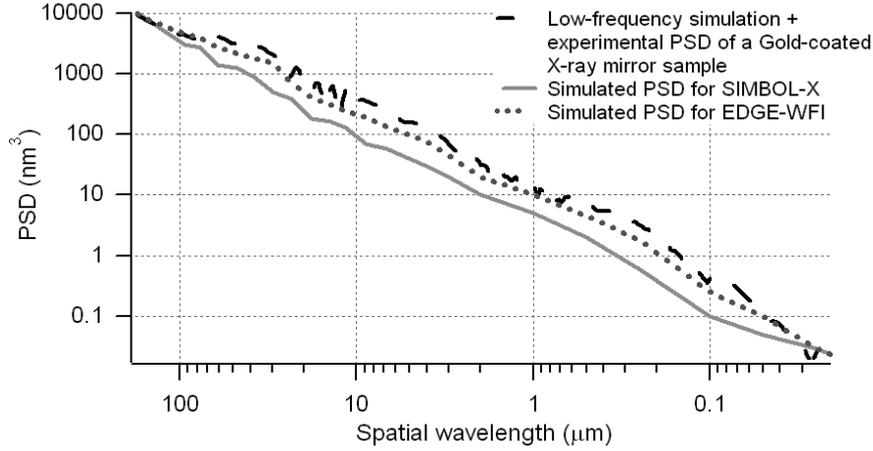

Fig. 3: Comparison of some possible PSD tolerances for the roughness of EDGE-WFI mirrors. The solid and the dashed line are the same of Fig. 1 (the range 100- 0.02 μm has been magnified). A suitable PSD for EDGE-WFI is represented by the dotted line. The tolerance is less exacting than for the case of SIMBOL-X (see Sect. 3).

Tab. 3: rms roughness, in selected wavelength windows, for some plotted PSDs in Fig. 3

| PSD | 1 mm > $l$ > 100 μm | 100 μm > $l$ > 10 μm | 10 μm > $l$ > 1 μm | $l$ < 1 μm |
|---|---|---|---|---|
| **Initial PSD** (squares + dashed line) | 10.2 Å | 3.6 Å | 2.5 Å | 1.8 Å |
| **Improved PSD** (dotted line) | 9.4 Å | 3.0 Å | 1.9 Å | 1.4 Å |

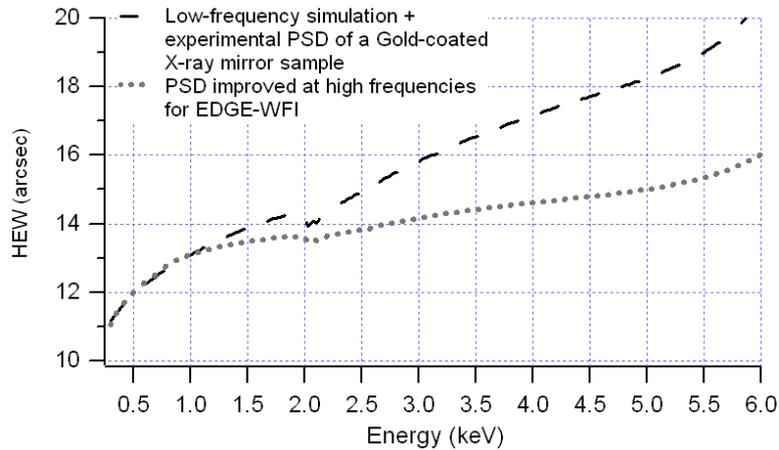

Fig. 4: HEW trends for the EDGE-WFI optical module, as computed from two PSD in Fig. 3. The dashed line is computed from the experimental PSD of the X-ray mirror sample referred to in Sect. 3 (Fig. 3, dashed line). The other HEW trend (dots) is derived from the improved PSD (dots in Fig. 3).

## 5. HEW SIMULATIONS FOR THE XEUS X-RAY TELESCOPE

The XEUS X-ray telescope is based on an alternative baseline. Its very large optics diameter (~ 4 m) and the consequent very large focal length (35 m) will force the adoption of the formation flight configuration. High Performance Pore Optics (HPO)[15,16] are a technology being developed to manufacture very large lightweight X-ray optics, so it can be suitable to manufacture the XEUS optics. They will comprise several optic petals, assembled from ribbed and bended Silicon plates, that will provide the optically-polished surface to reflect and focus X-rays in the XEUS focal plane.

*Excellent smoothness* is required to Silicon plates for HPO in order to keep the X-ray scattering at a level as low as possible. At the same time, a mass-production process is needed to produce the very large amount of Silicon plates to be assembled in the XEUS optics. Several factories produce Silicon wafers, optically polished down to a few angström rms. Here we discuss the particular case of Silicon wafers produced by MEMC Electronics Materials Inc. (St. Peters, MO, USA). The surface roughness of Silicon plates of this kind has been extensively characterized[18], and its PSD has been found to approximate very well a power-law:

$$P(f) \approx \frac{2 \; nm^3}{f(\mu m^{-1})^{1.7 \div 1.8}} \qquad (10)$$

The PSD of a MEMC Silicon wafer is plotted in Fig. 5 (green triangles). For comparison, also the measured PSD of the X-ray mirror sample reported in Sect. 3 is overplotted (dashed line). Even if the spectral index *n = 1.8* is quite steep and the PSD is almost everywhere lower that of the Gold-coated mirror sample, the quite large incidence angles - when compared to SIMBOL-X, see Tab. 1 - enhance the X-ray scattering influence on the HEW, so the HEW increases significantly with the photon energy. The resulting HEW trend (as calculated via the Eqs. 6,7,8, assuming $H_0$ = 5 arcsec) for the XEUS mirror module is plotted in Fig. 6 (triangles).

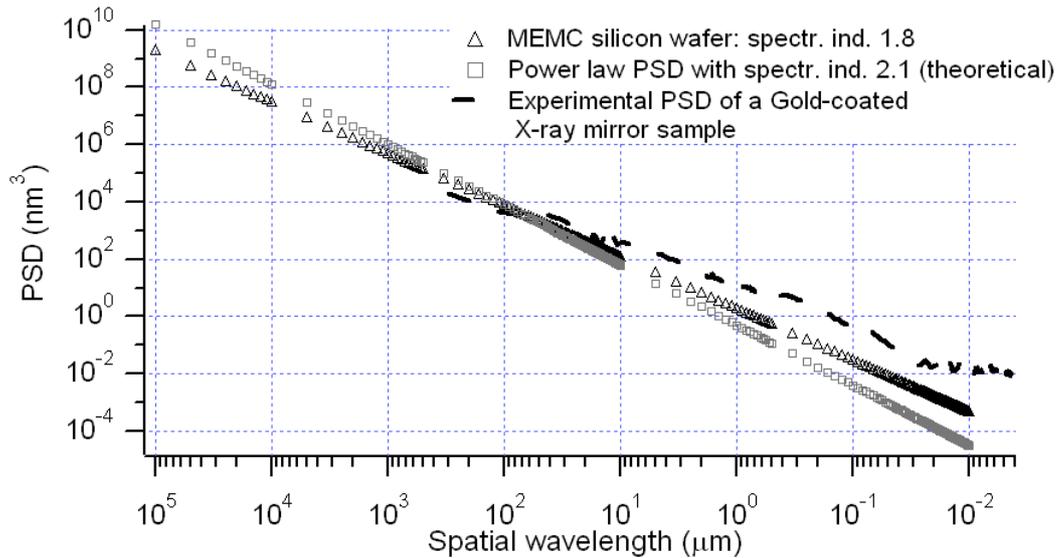

Fig. 5: the PSD of a commercial Silicon wafer (green triangles) and a *theoretical* PSD with spectral index 2.1 (squares).

To show the relevant influence of the spectral index of the PSD on the imaging degradation at high energies, a *simulated* PSD model has also been considered, a power-law with similar PSD values around 100 μm, but a spectral index *n* = 2.1 (see Fig. 5, squares). The PSD with *n* = 2.1 exhibit a smaller high-frequency content, while it is rougher at low frequencies. However, the damping of high frequencies causes a relevant reduction of the HEW at high energies. The HEW, as derived from this second PSD, is overplotted in Fig. 6: the HEW increase with the energy is much less marked (30 arcsec vs. 45 at 40 keV), even though at low energies the HEW turns out to be slightly larger, owing to the higher PSD values at low frequencies. This example highlights the benefits on the angular resolution deriving from a steeply decreasing PSD. Future development of substrates for X-ray mirrors should be aimed at a high-frequency cut-off as sharp as possible in the PSD. The rms values of the PSD with *n* = 2.1 in particular spectral windows are reported in Tab. 4 and compared with the corresponding values for the PSDs discussed in the previous sections.

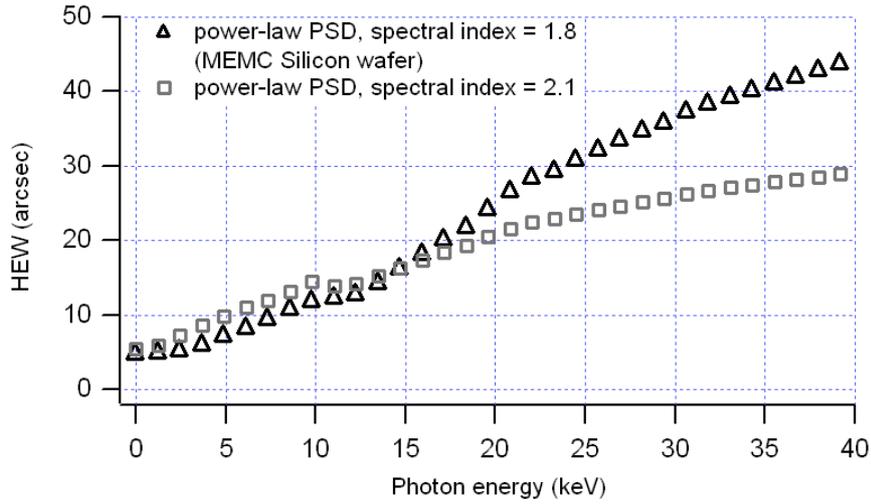

Fig. 6: the HEW for the XEUS optical module up to 40 keV, as derived from the PSDs in Fig. 5. The steeper PSD (squares) exhibits a smaller increase with the photon energy.

## 6. CONCLUSIONS AND FINAL REMARKS

The evaluation of the X-ray scattering impact in X-ray telescopes is a very important issue, because the XRS can seriously degrade the telescope angular resolution at high photon energies. To this end, a roughness tolerance should be formulated with reference to a specific requirement of HEW as a function of the photon energy. In particular, for a given mirror shell, *a mirror surface PSD can be immediately turned into a HEW function using the analytical treatment reported in the Sect. 2, and vice versa*. This makes in turn possible the evaluation of the HEW for all the optical module. The application of this method *allowed establishing some possible PSD tolerances for the surfaces of SIMBOL-X, EDGE-WFI, XEUS mirrors*, starting from reasonable HEW trends in the respective energy bands. Some rms values derived from the PSD requirements, and integrated in selected spectral ranges, are reported in Tab. 4. The practical realization of X-ray mirrors fulfilling the listed requirements, as well as a detailed study of PSD tolerances from the angular resolution requirements, would constitute a relevant leap forward in the development of optics for future X-ray telescopes with a very good imaging quality.

Tab. 4: possible roughness requirements for the optical modules under test.

|  | 1 mm > $\ell$ > 100 μm | 100 μm > $\ell$ > 10 μm | 10 μm > $\ell$ > 1 μm | $\ell$ < 1 μm |
|---|---|---|---|---|
| **SIMBOL-X** | 9.3 Å | 2.2 Å | 1.3 Å | 1.0 Å |
| **EDGE** | 9.4 Å | 3.0 Å | 1.9 Å | 1.4 Å |
| **XEUS** | 9.1 Å | 2.6 Å | 0.7 Å | 0.4 Å |

## APPENDIX A: THE HEW OF SEVERAL CONFOCAL SHELLS

Have *M* nested confocal X-ray mirror shells, with effective area $A_k$ (k =1…M). Each *k*-th mirror shell is characterized by its own PSF $S_k(\vartheta)$, that can be used to compute each encircled energy $EE_k(\vartheta)$. Each *k*-th Half-Energy-Radius $\Omega_k$ (in this case we consider both contribution of figure deformations and X-ray scattering, but we neglect assembly and alignment defects among mirrors), is defined by the relation

$$EE_k(\Omega_k) = \frac{1}{2}. \tag{11}$$

We evaluate now the HEW of the assembly formed by *M* mirror shells. We suppose the $S_k(\vartheta)$ functions to be normalized to 1 (i.e., the Encircled Energy approaches asymptotically 1 for large angles). The total, normalized PSF $S_T(\vartheta)$ is the superposition on the focal plane of all the normalized PSF $S_k(\vartheta)$, weighted over the respective effective areas $A_k$:

$$S_T(\vartheta) = \frac{\sum_1^M A_k S_k(\vartheta)}{\sum_1^M A_k}. \tag{12}$$

If we integrate the Eq. 12 over the angle $\vartheta$, we find that the same average holds for the Encircled Energy functions: we write it in this form:

$$EE_T(\vartheta) \sum_1^M A_k = \sum_1^M A_k EE_k(\vartheta). \tag{13}$$

The half-energy radius of the whole mirror shell assembly $\Omega_T$ fulfills the condition $EE_T(\Omega_T) = \frac{1}{2}$. In general, $\Omega_T \neq \Omega_k$ at least for some *k* (otherwise, the Eq. 13 would become a trivial identity at $\vartheta = \Omega_T$). It is reasonable, indeed, to suppose that $\Omega_T$ does not differ very much from each $\Omega_k$. Therefore, we assume a linear expansion of each $EE_k$ around $\Omega_k$,

$$EE_k(\Omega_T) \approx EE_k(\Omega_k) + \left.\frac{d(EE_k)}{d\vartheta}\right|_{\Omega_k}(\Omega_T - \Omega_k), \tag{14}$$

if we evaluate the Eq. 13 at $\Omega_T$, and substitute the Eq. 14 in the sum in Eq. 13 for all k, we obtain after some algebra

$$\sum_1^M A_k \left.\frac{d(EE_k)}{d\vartheta}\right|_{\Omega_k}(\Omega_T - \Omega_k) \approx 0, \tag{15}$$

where we canceled out all the other terms using the Eq. 11. Now, the derivatives of the Encircled Energy functions are simply $S_k(\Omega_k)$. Isolating $\Omega_T$ yields the simple result:

$$\Omega_T \approx \frac{\sum_1^M A_k S_k(\Omega_k) \cdot \Omega_k}{\sum_1^M A_k S_k(\Omega_k)} \tag{16}$$

finally, multiplying by 2 the Eq. 16, we derive the formula for the total $HEW_T$:

$$HEW_T \approx \frac{\sum_1^M A_k S_k(\Omega_k) \cdot HEW_k}{\sum_1^M A_k S_k(\Omega_k)}. \tag{17}$$

In other words, the resulting HEW is the weighted average of the HEWs of the individual shells: the weights are the effective areas $A_k$, multiplied by the respective normalized PSF $S_k$ at the respective half-energy radius $\Omega_k$.

## ACKNOWLEDGMENTS

The author is grateful to G. Pareschi, O. Citterio, V. Cotroneo, S. Basso, P. Conconi, F. Mazzoleni (*INAF-OAB*) for useful discussions and kind collaboration. He also acknowledges *ASI* (the Italian Space Agency), *INAF* (the National Institute for Astrophysics), *MUR* (the Italian Ministry for Universities) for the grants funding his fellowship.